\documentclass[conference]{IEEEtran}
\IEEEoverridecommandlockouts
\usepackage{cite}
\usepackage{amsmath,amssymb,amsfonts}
\usepackage{algorithmic,algorithm}

\usepackage{graphicx}
\usepackage{textcomp}
\usepackage{xcolor}

\newtheorem{theorem}{Theorem}

\def\BibTeX{{\rm B\kern-.05em{\sc i\kern-.025em b}\kern-.08em
		T\kern-.1667em\lower.7ex\hbox{E}\kern-.125emX}}

\begin{document}
\title{Security and Privacy Enhancement  for Outsourced Biometric Identification}
\author{Kai Zhou, Jian Ren and Tongtong Li\\
	Department of ECE, Michigan State University, East Lansing, MI 48824-1226\\
	Email: \{zhoukai, renjian, tongli\}@msu.edu}
\maketitle

\begin{abstract}
A lot of research has been focused on secure outsourcing of biometric identification in the context of cloud computing. In such schemes, both the encrypted biometric database and the identification process  are outsourced to the cloud. The ultimate goal is to protect the security and privacy of the biometric database and the query templates. Security analysis shows that previous schemes suffer from the enrolment attack and unnecessarily expose more information than needed. In this paper, we propose a new secure outsourcing scheme aims at enhancing the security from these two aspects. First, besides all the attacks discussed in previous schemes, our proposed scheme is also secure against the enrolment attack. Second, we model the identification process as a fixed radius similarity query problem instead of the kNN search problem. Such a modelling is able to reduce the exposed information thus enhancing the privacy of the biometric database. Our comprehensive security and complexity analysis show that our scheme is able to enhance the security and privacy of the biometric database and query templates while maintaining the same computational savings from outsourcing.	
\end{abstract}

\section{Introduction}
Remote Storage and computation outsourcing are two integer services provided by cloud computing. Data owners such as individuals or organizational administrators are able to outsource to the cloud their private data for storage as well as some computational intensive tasks for computation. Various works \cite{zhou2016ABE,zhou2016linsos,zhou2017expsos} have been devoted to securing the outsource process, i.e., ensuring the security of the private data while still enjoying the convenience provided by cloud computing.

Among various applications, outsourced biometric identification is of special interest. This is because, on one hand, the biometric database itself is of huge size thus making cloud storage an appealing solution. On the other hand, the identification process is computationally expensive due to the large database size. As a result, several schemes have been proposed to outsource biometric identification to the cloud. The ultimate goal of these schemes is to protect the security and privacy of the biometric templates in the database as well as the query templates under different attacks.

Securing outsourcing of biometric identification has attracted much research effort. In \cite{blanton2012secure}, the authors proposed two different schemes where both single-server model and multiple-servers model are considered. However, the scheme for the single-server model still has prohibitive computational overhead for large databases thus making it less practical. In the multiple-servers model, it is required that the severs would not collude, otherwise the security of the biometric templates is compromised. Following works \cite{chun2014outsourceable,elmehdwi2014secure} considered two non-colluding servers thus suffering from the same drawbacks. In \cite{wong2009secure}, a privacy-preserving biometric identification scheme was proposed for the single-server model. However, such a scheme is not secure under an active attack where the cloud is able to collude with query users. Most recently, the authors in \cite{wang2015cloudbi} proposed a secure outsourcing scheme that can defend against three different attacks as defined in \cite{wang2015cloudbi}, which include the attack where cloud and users are allowed to collude. The basic idea is to model the identification process as a Nearest Neighbor (NN)search problem. That is, given an encrypted query template, the nearest neighbor in the database is identified and returned to the data owner, who can later decide whether these two templates belong to the same individual.

Although the scheme in \cite{wang2015cloudbi} is efficient and can defend against relatively severe attacks as to date, recent analysis \cite{hahn2016poster} shows that it still suffers the enrollment attack. That is, if the cloud is allowed to inject selected templates into the database, then the cloud is able to recover the query templates based on the intermediate computation results. Also, we point out that modelling the identification process as kNN search problem has inherent limitations in terms of template privacy. To be more specific, in \cite{wang2015cloudbi}, given two encrypted templates $\mathbf{x}$ and $\mathbf{y}$ and an encrypted query template $\mathbf{z}$, the cloud can determine which template ($\mathbf{x}$ or $\mathbf{y}$) is closer to $\mathbf{z}$. Repeating this process, the cloud is able to identify the closet template to $\mathbf{z}$. As a result, the relative distance information among the templates in the database is inevitably exposed. 

To deal with the above two security issues, we propose a new secure outsourcing schemes for biometric identification. Especially, our proposed scheme can defend against all the attacks defined in \cite{wang2015cloudbi} as well as the enrollment attack. Moreover, we model the identification process as a fixed-radius similarity query problem rather than kNN search problem. That is, our scheme only enables the cloud to identify the templates within a fixed radius of the query template. No more information about the distance is revealed. In this way, less information is exposed compared to the scheme in \cite{wang2015cloudbi}, thus enhancing the privacy of the biometric database.

The rest of the paper is organized as follows. We introduce the system model and threat model in Section \ref{sec-model}. Then the secure outsourcing scheme is proposed in Section \ref{sec-scheme}. The security and complexity analysis is given in Section \ref{sec-analysis}. In Section \ref{sec-numeric}, we present some numeric results showing the efficiency of our proposed scheme. At last, we conclude in Section \ref{sec-conclusion}.

\section{Problem Formulation}\label{sec-model} 

\subsection{System model}

We consider a system consisting of three parties: a data owner, a set of end-users and a cloud service provider. The data owner has a biometric database composed of a collection of users' biometric templates, where each template can be represented by an $n$-dimensional vector $\mathbf{T} = (t_1,t_2,\cdots,t_n)$. Each template is registered by an end-user during an enrolment stage.  During a preparation stage, the data owner will pre-process the templates and outsource them to the cloud. Later, in an identification stage, an end-user will submit her identification request composed of a query template $\mathbf{T}_j$ to the data owner. The data owner will generate a token for each specific query template $\mathbf{T}_j$ and submit the token to the cloud. Then, the cloud is responsible for identifying the template $\mathbf{T}_i$ in the database such that $\mathsf{dist}(\mathbf{T}_i,\mathbf{T}_j) < \theta$, where $\mathsf{dist}()$ is a distance measurement function and $\theta$ is a pre-defined threshold. In words, the cloud will identify and return the template(s) whose distance from the query template is within a threshold.

\subsection{Threat model}

In this paper, we consider a \emph{semi-malicious} cloud model where the cloud will follow the protocols but is allowed to collude with some malicious end-users. To be specific, the collusion happens when some malicious end-users will submit query templates to the data owner and choose to share the templates with the cloud. As a result, the cloud is able to learn pairs of template and its encrypted form.

In summary, depending on the different capabilities of the  adversaries, we propose two attack models as follows.
\begin{enumerate}
\item Passive Attack: the cloud is able to know the encrypted templates $\mathsf{Enc}(\mathbf{T}_i)$, for $i=1,2,\cdots,m$, where $\mathsf{Enc}(\mathbf{T}_i)$ is the encrypted form of $\mathbf{T}_i$. Also, the cloud is able to observe a series of $w$ encrypted queries $\mathsf{Enc}(\mathbf{T}_j)$, $j=1,2,\dots,w$. However, the service provider does not know the underlying templates $\mathbf{T}_j$ in plaintext.

\item Active Attack: besides the encrypted templates $\mathsf{Enc}(\mathbf{T}_i)$, the cloud is able to observe a series of $w$ encrypted queries $\mathsf{Enc}(\mathbf{T}_j)$ as well as the corresponding plaintext $\mathbf{T}_j$, $j=1,2,\dots,w$. As mentioned earlier, such an attack can happen when the cloud collude with malicious end-users.
\end{enumerate}

Besides the above two attacks, we also allow the enrollment attack as considered in \cite{hahn2016poster}.

\begin{enumerate}
\item[3)] Enrollment Attack: the cloud is able to inject templates in the enrollment stage. That is, the cloud is able to have a series of $v$ encrypted templates $\mathsf{Enc}(\mathbf{T}_i)$ as well as the corresponding plaintext $\mathbf{T}_i$, $i=1,2,\dots,v$.
\end{enumerate}

Informally, the security requirement of the outsourced biometric identification against
the above three attacks is that the cloud is not able to learn more
information about the templates than what is allowed through the identification
process. That is, the cloud is only able to decide whether
the distance between two templates is within a threshold or not. It
is not feasible for the service provider to derive any key information
about the enrolled templates and the query templates.

\section{Secure Outsourcing of Biometric Identification}\label{sec-scheme}

\subsection{Basic Framework}
Intuitively, our proposed secure outsourcing scheme is consist of four phases. In the first phase, the data owner will generate the system parameters and a transformation key. Then, for each biometric template in the database, it is transformed to an encrypted form using the transformation key. The transformed database is then outsourced to the cloud. In the query phase, for every submitted query template, the data owner will generate a token using the same transformation key. At last, in the identification phase, the cloud will identify the template whose distance from the query template is within a pre-defined threshold.

Our proposed scheme is composed of the following five algorithms.
\begin{itemize}
\item $\mathsf{SetUp}()\to param$: the set up algorithm generate system parameters $param$.

\item $\mathsf{KeyGen}()\to sk$: the key generation algorithm will generate transformation key $sk$.

\item $\mathsf{Transform}(sk,\mathbf{x})\to C_{\mathbf{x}}$: given a vector $\mathbf{x}$ and transformation key $sk$, the transformation algorithm will transform  $\mathbf{x}$ into a disguised form $C_{\mathbf{x}}$.

\item $\mathsf{TokenGen}(sk,\mathbf{y})\to T_{\mathbf{y}}$: given a vector $\mathbf{y}$ and transformation key $sk$, the token generation algorithm will generate a token $T_{\mathbf{y}}$ for $\mathbf{y}$.

\item $\mathsf{Evaluate}(C_{\mathbf{x}},T_{\mathbf{y}})\to \Lambda=\{0,1\}$: given the transformed vector $C_{\mathbf{x}}$ and token $T_{\mathbf{y}}$, the evaluation algorithm will output a result $\Lambda$ satisfying
\[
\Lambda=\begin{cases}
1, & \mathsf{dist}(\mathbf{x}, \mathbf{y}) \leq\theta\\
0, & \text{otherwise},
\end{cases}
\]
where $\mathsf{dist}(\mathbf{x}, \mathbf{y})$ is the distance between $\mathbf{x}$ and $\mathbf{y}$ and $\theta$ is a pre-defined threshold.
\end{itemize}

\subsection{Secure Transformation and Evaluation}

The essential part in our proposed scheme is the secure transformation and evaluation process. In a high level view, the enrolled templates are transformed through the $\mathsf{Transform}$ function and the $\mathsf{TokenGen}$ function will generate a token by transforming the query template. It is critical that given the transformed templates, it it computationally infeasible to recover the original vector. However, the $\mathsf{Evaluate}$ function is able to reveal some information of two templates. That is, whether the distance between the two templates is within a threshold or not.

Our transformation process is similar to the techniques utilized in \cite{wang2015cloudbi}. However, the computational models as well as the security requirements are fundamentally different. We now give some intuition about our transformation and evaluation process. The detailed construction is presented in Protocol \ref{alg:outsourcing}. Given a vector $\mathbf{x}$, we first extend it to $\mathbf{x}'$ by inserting the threshold $\theta$ and some random numbers. Then $\mathbf{x}'$ is transformed to a matrix from and is disguised by multiplying it with random matrix.  Denote this disguised form as $C_x$. Given a query vector $\mathbf{y}$, the $\mathsf{TokenGen}$ function will transform $\mathbf{y}$ into a disguised form $C_y$. The $\mathsf{Evaluate}$ function takes in $C_x$ and $C_y$ as input and outputs $\alpha \beta (\mathbf{x} \circ \mathbf{y} - \theta)$, where $\alpha$ and $\beta$ are one-time random positive numbers associated with $\mathbf{y}$ and $\mathbf{x}$, respectively. By comparing $\alpha \beta (\mathbf{x} \circ \mathbf{y} - \theta)$ with $0$, the cloud is able to determine whether the inner product of $\mathbf{x}$ and $\mathbf{y}$ is within the threshold $\theta$ or not. We note that since $\alpha$ and $\beta$ are one-time random numbers and are different for each template, the exact value of $(\mathbf{x} \circ \mathbf{y}-\theta)$ can be concealed from the final result. We note that the inner product $\mathbf{x} \circ \mathbf{y}$ is flexible to express different distance metrics between $\mathbf{x}$ and $\mathbf{y}$.

\subsection{The Proposed Scheme}

In this section, we give the detailed implementation of our secure outsourcing scheme in Protocol \ref{alg:outsourcing}. In the protocol, the result $I$ in function $\mathsf{Evaluate}$ is equal to $(\mathsf{dist}_E(\mathbf{x}, \mathbf{y})-\theta)$, where $\mathsf{dist}_E(\mathbf{x}, \mathbf{y})$ is the Euclidean distance  between $\mathbf{x}$ and $\mathbf{y}$. We use the Euclidean distance as an example to measure the similarity between two templates. However, it is easy to design the vectors $\mathbf{x}$ and $\mathbf{y}$ such that the $\mathsf{Evaluate}$ function will give other distances such as the Hamming distance. The correctness of our proposed scheme is shown in Theorem \ref{them:correct}.
\begin{algorithm}[tbh]
	\floatname{algorithm}{Protocol}
	\caption{Secure Outsourcing of Biometric Identification}{\label{alg:outsourcing}}
	\smallskip 
	\textbf{Input:} $\mathbf{x} =\{x_1,\dots,x_n\},\mathbf{y} =\{y_1,\dots,y_n\},\theta$.\\
	\textbf{Output:} $\Lambda=\{0,1\}$.
	
	\smallskip 
	$\mathsf{Setup()}\rightarrow param$:
	\begin{algorithmic}[1] 
		\STATE Data owner sets $param = \{ n,\theta\}$., where $n$ is the dimension of templates and $\theta$ is a pre-defined threshold.
	\end{algorithmic} 
	
	\smallskip 
	$\mathsf{KeyGen}(\lambda)\rightarrow sk $:
	\begin{algorithmic}[1] 
		\STATE Randomly generates two matrices $M_1$ and $M_2$ with dimension $(n+5)\times (n+5)$ and a permutation $\pi: \mathbb{R}^{n+5} \rightarrow \mathbb{R}^{n+5}$.
		\STATE Set  $sk =\{M_1,M_2,M_1^{-1},M_2^{-1},\pi\}$
	\end{algorithmic} 
	
	\smallskip
	$\mathsf{Transform}(sk,\mathbf{x})\rightarrow C_{\mathbf{x}}$:
	\begin{algorithmic}[1]
		\STATE Generate  random numbers $\beta$ and $r_x$.
		\STATE (\textbf{Extend})  Extend $\mathbf{x}$ to an $(n+5)$-dimensional vector $\mathbf{x}' = (2\beta x_1,2\beta x_2,\dots,2\beta x_n,-\beta \sum\limits_{i=1}^nx_i^2,\beta ,\beta \theta^2,r_x,0)$.
		\STATE (\textbf{Permute}) Permute  $\mathbf{x}'$ to obtain $\mathbf{x}'' = \pi(\mathbf{x}')$.
		\STATE Transform $\mathbf{x}''$ a diagonal matrices $X$ with  $\mathbf{x}''$ being the diagonal.
		\STATE Generate a random  $(n+5)\times (n+5)$ lower triangular matrix $S_x$ with the diagonal entries fixed as $1$. Compute $C_x = M_1 S_x X M_2$.
	\end{algorithmic} 
	
	\smallskip
	$\mathsf{TokenGen}(sk,\mathbf{y})\rightarrow T_{\mathbf{y}}$:
	\begin{algorithmic}[1] 
		\STATE On receiving a query template $\mathbf{y}$, data owner generates random numbers $r_y$ and $\alpha$.
		\STATE (\textbf{Extend}) Extend $\mathbf{y}$ to an $(n+5)$-dimensional vector $\mathbf{y}' = (2\alpha y_1, 2\alpha y_2,\dots,2\alpha y_n,\alpha,-\alpha \sum\limits_{i=1}^ny_i^2,\alpha, 0,r_y)$.
		\STATE (\textbf{Permute}) Permute $\mathbf{y}'$ to obtain $\mathbf{y}'' = \pi(\mathbf{y}')$.
		\STATE Transform $\mathbf{y}''$ to a diagonal matrix $Y$ with diagonal being $\mathsf{y}''$.
		\STATE Generate a random $(n+5)\times (n+5)$ lower triangular matrix $S_y$ with the diagonal entries fixed as $1$. Compute $C_y = M_2^{-1} Y S_y  M_1^{-1}$
		\STATE Send the token $C_y$ to the cloud.
	\end{algorithmic} 
	
	\smallskip
	$\mathsf{Evaluate}(C_{\mathbf{x}},C_{\mathbf{y}})\rightarrow\Lambda$:
	\begin{algorithmic}[1] 
		\STATE For every transformed template $C_x$ in the database, the cloud computes $I = \mathsf{Tr}(C_x C_y)$, where $\mathsf{Tr}(\cdot)$ is the trace of a matrix.
		\STATE Cloud sets $\Lambda = 1$ if $I \leq 0$, which means that the template $\mathbf{x}$ is identified; otherwise set $\Lambda = 0$.
	\end{algorithmic} 
\end{algorithm}

\begin{theorem}\label{them:correct}
	The proposed outsourcing scheme in Protocol \ref{alg:outsourcing} is correct. That is, $I=\alpha\beta(\mathsf{dist}_E(\mathbf{x}, \mathbf{y})-\theta)$, where $\mathsf{dist}_E(\mathbf{x}, \mathbf{y})$ is the Euclidean distance  between $\mathbf{x}$ and $\mathbf{y}$.
\end{theorem}

\begin{IEEEproof}
	For a square matrix $Y$, the trace $\mathsf{Tr}(Y)$ is defined as the sum of
	the diagonal entries of $Y$. Given an invertible matrix $M_1$ of the same size, the transformation $M_1YM_1^{-1}$ is called similarity transformation of $Y$. From linear algebra, we know the trace of a square matrix remains unchanged under similarity transformation. That is, $\mathsf{Tr}(Y)=\mathsf{Tr}(M_1YM_1^{-1})$. Then we have
	\[
	I=\mathsf{Tr}(T_1)+\mathsf{Tr}(T_2)=\mathsf{Tr}(S_pPYS_y)+\mathsf{Tr}(S_qQYS_y).
	\]
	Since $S_y$, $S_p$ and $S_q$ are selected as lower triangular matrices, where all the diagonal entries are set to $1$, the diagonal entries of $S_pP$, $S_qQ$ and $YS_y$ are all the same as those of $P$, $Q$ and $Y$. Thus we have
	\[
	I=\mathsf{Tr}(PY)+\mathsf{Tr}(QY).
	\]
	Since $P$,$Q$ and $Y$ are diagonal matrices, we have $\mathsf{Tr}(PY) =\mathbf{p}\circ\mathbf{y}'$ and $\mathsf{Tr}(QY) =\mathbf{q}\circ\mathbf{y}'$. Thus
	\[
	I=\mathbf{p}\circ\mathbf{y}'+\mathbf{q}\circ\mathbf{y}'=\mathbf{x}'\circ\mathbf{y}'=\alpha\beta(\mathsf{dist}_E(\mathbf{x}, \mathbf{y})-\theta).
	\]

	\vskip-1.5\baselineskip

\end{IEEEproof}

\section{Security and Complexity Analysis}\label{sec-analysis}

\subsection{Security against Active Attack}

We focus on the security of our proposed scheme under active attack since it implies the security under passive attack. We also utilize $\mathsf{TokenGen}$ function as the representative since the transformation process in $\mathsf{Transform}$ is similar. The basic idea is to show that an adversary cannot differentiate two transformed templates obtained from the $\mathsf{TokenGen}$ function. Thus, the adversary cannot learn key information from the disguised form of templates. We have the following theorem.  

\begin{theorem}\label{thm:active}
The proposed outsourcing scheme is secure against active attack, that is the cloud cannot derive key information from transformed templates.
\end{theorem}

\begin{IEEEproof}
Consider the transformation of vector $\mathbf{y}$, where $\mathbf{y}=(y_1,\dots,y_n)$. The vector $\mathbf{y}$ is first extended to $\mathbf{y}'=(2\alpha y_1, 2\alpha y_2,\dots,2\alpha y_n,\alpha,-\alpha \sum\limits_{i=1}^ny_i^2,\alpha, r)$. The vector $\mathbf{y}'$
is then extended to a diagonal matrix $Y$.
Then, it is transformed to  $C_y=M_2^{-1}YS_yM_1^{-1}$, where $S_y$ is a random lower triangular matrix. We note that
the product of $Y$ and $S_y$ will produce a lower triangular
matrix denoted as $G_y$. Now we
focus on the product $C_y=M_2^{-1}G_yM_1^{-1}$. 

Denote the entries in $M_2^{-1}$ and $M_1^{-1}$ as $a_{ij}$
and $b_{ij}$, respectively, where $i,j=1,2,\dots,n+5$. For matrix
$G_y$, denote its non-zero entries in the lower triangular part
as $s_{ij}$, where $i>j$ and $i,j=1,2,\dots,n+5$. Then, by law
of matrix multiplication, each entry $c_{ij}$ in $C_y$ can be
written in the form of 
\begin{eqnarray}
c_{ij} & = & \sum f_{ij}^1(a_{ij},b_{ij})m_i +f_{ij}^2(a_{ij},b_{ij})\alpha\nonumber\\
&  & +f_{ij}^{3}(a_{ij},b_{ij})r +f_{ij}^{4}(a_{ij},b_{ij},s_{ij}),
\label{eq:entry}
\end{eqnarray}
where $f_{ij}^{t}$, $t=1,2,3,4$ are polynomials. Equation (\ref{eq:entry})
is obtained by summing up each terms of $m_i$, $\alpha$ and $r$,
respectively.

In the transformation process, $a_{ij}$
and $b_{ij}$ are fixed. $\alpha$,$r$ and $s_{ij}$ are one-time random
numbers. $m_i$ are chosen and can be controlled by the adversary. However,
since $\alpha$,$r$ and $s_{ij}$ are one-time random numbers, the polynomials
$f_{ij}^2(a_{ij},b_{ij})\alpha$, $f_{ij}^{3}(a_{ij},b_{ij})r$
and $f_{ij}^{4}(a_{ij},b_{ij},s_{ij})$ all looks random to the adversary.
As a result, the summation $c_{ij}$ is random.
This means that, for any two templates chosen by the adversary and
one transformed template, the adversary cannot distinguish which
template is actually transformed. As a result, the adversary cannot derive key information from transformed templates.

\vskip-1.0\baselineskip

\end{IEEEproof}

The other important aspect of security is to what extent the $\mathsf{Evaluate}$ function can reveal information of the templates. It is clear that $\mathsf{Evaluate}$ will the distance information which is necessary for identification.  However, we note that  every vector $\mathbf{y}$ is associated with a one-time independent
random number $\alpha$ and every vector $\mathbf{x}$ is associated
with a one-time random number $\beta$. As a result, in the active
attack, what an adversary can observe through $\mathsf{Evaluate}$ function is a series
of results $I_i=\alpha\beta_i(\mathsf{dist}_E(\mathbf{x},\mathbf{y})-\theta)$.
Since $\beta_i$ are selected independently, the final results
$I_i$ only reveals whether $\alpha\beta_i(\mathsf{dist}_E(\mathbf{x},\mathbf{y})-\theta)$
is positive or not. No more key information can be derive from $I_i$.

\subsection{Security against Enrolment Attack}

As mentioned earlier, an enrolment attack was proposed in \cite{hahn2016poster} making  the secure outsourcing scheme in \cite{wang2015cloudbi} vulnerable. In an enrolment attack, an adversary (i.e., the cloud) is able to inject known templates into the database. During evaluation,
the cloud is able to derive the following equation (i.e., Equation
(3) in \cite{hahn2016poster}):
\[
b_{ci}=\frac{(\mathsf{Tr}(Y_i^{'}B_{c}^{'})-\mathsf{Tr}(X_i'B_{c}^{'}))-(y_{i(n+1)}-x_{i(n+1)})}{y_{ii}-x_{ii}},
\]
where $b_{ci}$ is the $i$-th entry in a submitted query template
$\mathbf{b}_{c}$. Since $\mathsf{Tr}(Y_i^{'}B_{c}^{'})$ and $\mathsf{Tr}(X_i'B_{c}^{'})$
are computable and $\mathbf{x}$ and $\mathbf{y}$ are selected by
the cloud, the cloud is able to recover $b_{ci}$. Repeating such
attack will finally recover the whole query template $\mathbf{b}_{c}$
as demonstrated in \cite{hahn2016poster}. 

We now show that our proposed  scheme is secure under the above enrolment attack.
The underlying reason that the scheme in \cite{wang2015cloudbi} cannot
defend such attack is that the evaluate function will cancel all the randomness (i.e., the random lower-triangular matrix) introduced in the encryption process.
In comparison, the evaluation function in our scheme will give $\alpha_{z}\beta_x(\mathsf{dist}^2(\mathbf{x},\mathbf{z})-\theta)$,
where $\alpha_{z}$ and $\beta_x$ are one-time random numbers associated
with the templates $\mathbf{z}$ and $\mathbf{x}$ respectively. As a result, Equation
(3) in \cite{hahn2016poster} is modified to 

\[
b_{ci}=\frac{(\mathsf{Tr}(PB_{c}^{'})-\mathsf{Tr}(QB_{c}^{'}))-(p_{n+1}-q_{n+1})}{\alpha_{c}\beta_x(p_n-q_n)}.
\]
Note that $\alpha_{c}$ is a one-time random number associated with
a query $b_{c}$ and $\beta_x$ is a one-time random number associated
with $\mathbf{x}$. Thus, although the adversary is able to insert
known templates into the database, it cannot derive $b_{ci}$ due
to the one-time randomness. In other words, our proposed outsourcing scheme is able to defend against the enrolment attack.

\subsection{The Effect of Randomness on Security\label{subsec:The-Effect-of}}

It is important to understand the effect of different randomness on
security. We briefly categorize the one-time randomness utilized by our scheme into three types.
\begin{itemize}
	\item Type I: \textit{result-disguising randomness}. In the \textbf{Extend}
	step in both $\mathsf{Transform}$ and $\mathsf{TokenGen}$, we
	use random $\beta$ and $\alpha$ respectively to multiple with each
	entry of $\mathbf{x}$ and $\mathbf{y}$. Since $\alpha$ and $\beta$
	will remain in the decryption result, we name it as result-disguising
	randomness.
	\item Type II: \textit{vector-extension randomness}. In $\mathsf{TokenGen}$,
	we extend the vector $\mathbf{y}$ and pad it with a random $r$.
	\item Type III: \textit{matrix-multiplication randomness}. In both $\mathsf{Transform}$
	and $\mathsf{TokenGen}$, we multiple the extended matrices ($P$,
	$Q$ and $Y$) with random matrices ($S_p$, $S_q$ and $S_y$).
\end{itemize}

The $\mathsf{Evaluate}$ function will calculate the trace
of the matrix (e.g., $C_pC_y$). We note that the function $\mathsf{Tr}(\cdot)$
will cancel Type II and Type III randomness. However, Type I randomness
will remain in the evaluation result.
This is important since it will only reveal partial information of
the plaintext, which is just sufficient for the purpose of biometric
authentication. Also, the underlying reason that the scheme in \cite{wang2015cloudbi} is vulnerable to enrollment attack is that it lacks Type I randomness.

\subsection{Complexity Analysis}
We focus on the complexity analysis of $\mathsf{TokenGen}$ and $\mathsf{Evaluate}$ since they are executed repeatedly in the identification process while $\mathsf{SetUp}$, $\mathsf{KeyGen}$ and  $\mathsf{Transform}$ are one-time processes. As shown in Protocol \ref{alg:outsourcing}, it is obvious that the computational bottleneck of $\mathsf{TokenGen}$ and $\mathsf{Evaluate}$ lies in matrix multiplication. Without loss of generality, we assume that the matrices involved in the computation all have the same dimension $n\times n$. 

The function $\mathsf{TokenGen}$ will take $3$ matrix multiplications. Since matrix multiplication generally has a complexity of $\mathcal{O}(n^3)$ without optimization, the complexity of $\mathsf{TokenGen}$ is also $\mathcal{O}(n^3)$. In the function $\mathsf{Evaluate}$, the trace of two matrices $C_p C_y$ and $C_q C_y$ need to be computed. We note that there is no need to calculate the matrix multiplication first. What needs to be computed are the main diagonals of the two matrices. Thus, $\mathsf{Evaluate}$ has a complexity of $\mathcal{O}(n^2)$.

In terms of communication overhead, we assume that each entry in
the matrix or vector has the same size $l$. For each template in the database, 
the data owner needs to outsource the encrypted template $C_x=\{C_p,C_q\}$
to the cloud. Thus the communication overhead
is $2n^2l$. Similarly, the communication overhead for each query template is $n^2l$. 

\section{Numeric Results}\label{sec-numeric}
The most important parameter that affects the performance of the identification process is the length of the vectors denoted as $n$. It will determine the execution time for both $\mathsf{Transform}$ and $\mathsf{Evaluate}$, which are executed frequently in the querying process. Another parameter is the size of the database $N$. However, since our identification algorithm is basically a linear scan of the database, it can be predicted that the time for identification is also linear to $N$. Thus, it is of more interest to measure the performance of $\mathsf{Transform}$ and $\mathsf{Evaluate}$ in terms of the dimension $n$.

\begin{figure}[tbh]
	\centering
	\includegraphics[width=0.5\textwidth]{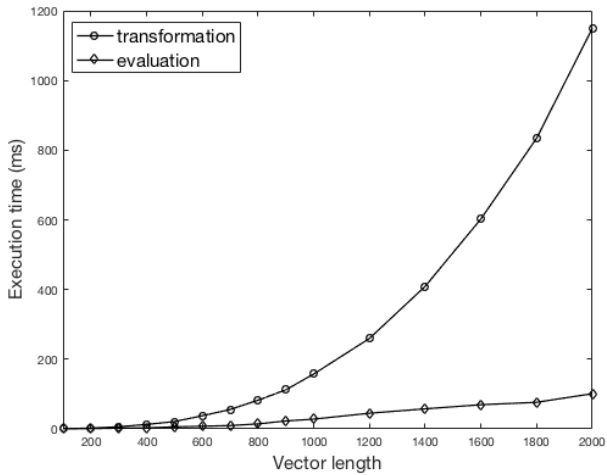}
	\caption{Template transformation and evaluation time for each template}
	\label{exeTime}
\end{figure}

The simulation is conducted in a personal computer with 1.6 GHz Intel Core i5 CPU, 4 GB RAM and macOS Version 10.12.6. The algorithm is implemented using the Armadillo C++ linear algebra library.  In the simulation, we let the length of the vector vary from 100 to 2000, which is able to cover the length of some typical biometric templates such as FingerCodes \cite{jain2000filterbank}. The execution time of $\mathsf{Transform}$ and $\mathsf{Evaluate}$ for each template is presented in Fig.\ref{exeTime}. We can see that both $\mathsf{TokenGen}$ and $\mathsf{Evaluate}$ are quite efficient. For example, it takes around 1 second to generate a token for a template with length $2000$, which is quite long in real applications.  The numeric results also correspond with the complexity analysis that $\mathsf{Transform}$ has $\mathcal{O}(n^3)$ complexity while $\mathsf{Evaluate}$ has $\mathcal{O}(n^2)$ complexity.

\section{Conclusion}\label{sec-conclusion}
In this paper, we proposed a new secure outsourcing scheme for biometric identification aiming at enhancing the security and privacy for the outsourced biometric database and the query templates. Specifically, our scheme is able to defend against the enrollment attack that makes previous schemes vulnerable. By modelling identification as fixed radius similarity search problem, our scheme exposes less information than previous schemes that based on kNN search problem. In summary, our comprehensive security and complexity analysis show that our scheme is able to enhance the security and privacy of the biometric database and query templates while maintaining the same computational savings from outsourcing.	
\bibliographystyle{plain}
\bibliography{citation}

\end{document}